# Internal inconsistencies and unresolved experimental evidence in a quantum neurosecurity framework

Grigorovich S., Bairachnaya M., Daroczi E., Marshall J., Giros B.
Douglas Mental Health Institute, McGill University, Montreal, QC, Canada

## Summary

Ben Othman and Ali introduce the Quantum-Entangled Neuro-Symbolic Swarm Federation (QENSSF), a framework intended for privacy-preserving, quantum-secure, AI-powered, multimodal healthcare data analysis [1]. They claim QENSSF is a first-in-class framework with provable privacy and security guarantees, support these with theorems and proofs, report experimental evaluation on IBM's quantum Eagle processor, report a real-world quantum key distribution (QKD) test, and claim use of real-world multimodal datasets (ADNI, UK Biobank, and MIMIC-IV). Our examination of quantitative support for these claims identified various concerns. Most importantly, the security criterion of a central guarantee (Theorem 1) cannot be satisfied given the risk formulation and relevant constraints, which, along with calculation errors in the proof, directly undermines the claim of provable security. Furthermore, insufficient reporting of biomedical datasets and unresolved discrepancies in the hardware evaluation limit verification of the experimental claims and overall reproducibility. QENSSF identifies a real need: ensuring privacy and security of healthcare data while patients benefit from AI-powered analysis. This need is particularly pressing for sensitive neural data, including the neuroimaging data available through sources such as ADNI. Privacy concerns also extend to wearable EEG data, for which reidentification risks can persist despite deidentification [2]. However, QENSSF's foundational inconsistencies need resolution before it could be considered in the neurosecurity domain.

## The stated security condition is incompatible with the risk formulation

The article defines the system risk as

$$R_{\text{sys}} = \lambda_C I_Q + \lambda_I P_{\text{viol}} + \lambda_A P_{\text{unavail}} + \mu_Q V_{\text{CHSH}},$$

with weights $\lambda_C = 0.55$, $\lambda_I = 0.25$, $\lambda_A = 0.15$, and $\mu_Q = 0.05$. This is the single-step expression used in the article's numerical demonstrations; Eq. (39) additionally sums over discrete time steps. According to Theorem 1, the framework is considered secure when $R_{\text{sys}} < 0.03$.

However, the stated security condition is incompatible with the authors' own definitions. The article identifies $V_{\text{CHSH}} > 2$ as confirmation of entanglement, with $V_{\text{CHSH}} < 2.8$ as an upper bound. But with the specified $\mu_Q = 0.05$ and $V_{\text{CHSH}} > 2$, the CHSH contribution alone is

$$\mu_Q V_{\text{CHSH}} > 0.05 \times 2 = 0.10 > 0.03.$$

Since the remaining terms are non-negative, no values of the other components can satisfy the security condition $R_{\text{sys}} < 0.03$ under the proposed definition. Even with zero quantum information leakage, integrity violations, and system unavailability, the CHSH contribution alone keeps the total risk above the specified threshold. The incompatibility follows from the risk formulation, independently of the theoretical or observed values of the remaining variables.

Furthermore, the inclusion of the CHSH score with a positive coefficient raises a conceptual inconsistency. The article treats CHSH inequality violation as evidence of non-local quantum correlations supporting the framework's security; yet increasing $V_{\text{CHSH}}$ increases the defined system risk. The property presented as desirable for quantum security contributes positively to a metric that must remain below a restrictive threshold.

The numerical demonstrations supporting Theorem 1 contain additional inconsistencies. In the first numerical substitution,

$$0.55 \times 10^{-14} + 0.25 \times 0.018 + 0.15 \times 2 \times 10^{-4} + 0.05 \times 2.8,$$

the article reports $R_{\text{sys}} = 0.02453 < 0.03$, while direct evaluation gives

$$R_{\text{sys}} = 5.5 \times 10^{-15} + 0.0045 + 0.00003 + 0.14 \approx 0.14453.$$

The value is almost five times the security threshold. As such, the intermediate terms reported in the article do not sum to the reported result. Thus, the security condition is not satisfied.

The discrepancy recurs in the subsequent simulation example. In the second calculation, using the parameter values reported from the simulation,

$$0.55 \times (1.2 \times 10^{-15}) + 0.25 \times 0.015 + 0.15 \times (9.8 \times 10^{-5}) + 0.05 \times 2.68,$$

the authors report a result of approximately $0.0249 < 0.03$. However, with direct substitution,

$$R_{\text{sys}} = 6.6 \times 10^{-16} + 0.00375 + 0.0000147 + 0.134 \approx 0.137765.$$

While the authors claim both results fall below the security threshold, correct evaluation shows that both exceed it. Thus, these examples do not support the claimed security criterion; rather, they appear to support it only because of incorrect arithmetic.

The corrected calculations illustrate the underlying incompatibility: with the stated constraints, the security condition cannot be satisfied in the CHSH-violating regime claimed for QENSSF.

Whether this reflects an error in the risk definition or its implementation requires clarification; therefore, Theorem 1, as currently formulated, does not establish the claimed information-theoretic security guarantee for QENSSF in the CHSH-violating operating regime.

## Hardware execution cannot be established by the reported methods

The abstract explicitly states that QENSSF was evaluated on IBM's 127-qubit Eagle processor. However, the article does not provide device-specific execution records, and how the authors define the extent of their hardware evaluation remains ambiguous.

The experimental section, titled "Simulation environment and network configuration", identifies only the Qiskit Aer simulator for circuit emulation, and the circuit section reports 10,000-shot Aer Monte-Carlo simulations. Hardware details such as physical backend name, job identifier, execution date, calibration record, transpiled circuit, qubit mapping, or raw hardware counts are not specified. To clarify whether "evaluation on Eagle" means physical execution or simulation, we examined the authors' subsequent QEDRA article [3]. It states that all experiments were performed in a hybrid simulation environment with its quantum layer simulated in Qiskit Aer using a model of Eagle, yet concludes the system was validated on Eagle processors. This suggests the authors may use "validated on Eagle" to include Eagle-parameterized Aer simulations, but the original article gives no definitive explanation. Simulations can be useful but do not support claims of hardware validation.

The claimed real-world QKD test raises a similar issue. The authors report transmitting 3,000 EEG samples over 10 km of fiber with a QBER of approximately 0.03, but omit photon-source type, wavelength, mean photon number, detector characteristics, measured channel loss, encoding arrangements, error-correction implementation, and privacy-amplification implementation. This is insufficient to reproduce a physical QKD experiment or determine the extent of simulation.

Moreover, the classical hardware description conflicts with the manufacturer's specifications. QENSSF describes NVIDIA A100 GPUs with 80 GB HBM3 and 1.4 TFLOPS FP64 performance. NVIDIA specifies the A100 80 GB with HBM2e, 9.7 TFLOPS conventional FP64 performance, and 19.5 TFLOPS FP64 Tensor Core performance [4]. This matters because throughput and energy results for a 128-GPU cluster are reported, and the inconsistency complicates independent interpretation and reproduction.

## The experimental datasets cannot be reconstructed

Reportedly, experimentation utilized five data modalities: wearable sensor, functional MRI, EEG, genomic, and clinical text data. The authors cite each modality directly (references 25–29 of the original article), and the abstract lists three public data resources: ADNI [5], UK Biobank [6], and MIMIC-IV [7], which contain relevant data (genomic, functional MRI, wrist accelerometry, and critical-care records). Access to the specific data used is critical for independently verifying QENSSF's multimodal integration.

However, references 25–29 do not identify or describe modality-specific datasets, nor do they refer to the listed public resources or document evidence of their use in the study. Their secondary citation chains do not recover the dataset sources with reported cohort sizes and institutional partitions. Therefore, it remains unclear whether the reported numbers of participating institutions represent actual sites or synthetic federated partitions of existing datasets. Although the article gives counts for all modalities from 20–50 institutions, it omits information required to establish the origin of these site-level partitions, such as dataset versions, record identifiers, selection criteria, or source-site variables. The authors' later QEDRA paper partitions MIMIC-IV records among simulated sites using a Dirichlet distribution [3]; the QENSSF manuscript reports no comparable procedure.

Finally, although the authors state the datasets are available upon reasonable request, our request has received no response. Clinical datasets are often access-restricted, but preprocessing, cohort-selection, and partitioning methods could be provided independently of protected information. Without data provenance, QENSSF's multimodal integration cannot be verified.

## Conclusion

The issues we raise concern the authors' central claims: the risk equation reveals that QENSSF cannot meet its own security threshold and is not provably secure as published; insufficient information is provided to establish the extent of the claimed physical IBM quantum hardware execution; and inadequate dataset reporting prevents independent reproduction of the multimodal evaluation. Presently, QENSSF requires a revised risk formulation, corrected security calculations, verifiable quantum-hardware execution records, complete experimental documentation, and traceable data sources with reproducible analysis procedures. These discrepancies go beyond reproducibility and raise fundamental doubts about whether the evidence presented supports QENSSF's principal security and experimental claims.

## Competing interests

The authors declare no competing interests.

## Data availability

No new experimental datasets were generated. The mathematical and methodological assessment is based on information reported in the published article and sources cited.